\documentclass[10pt]{article}
\usepackage{amssymb}
\usepackage{epsfig,graphicx}
\usepackage{enumerate}
\usepackage{url} 

\newcommand{\blind}{0}

\newcommand{\Prob}{{\rm I\hspace{-0.8mm}P}}
\newcommand{\E}{{\bf E}}
\newcommand{\RL}{{\rm I\hspace{-0.8mm}R}}

\newcommand{\bfP}{\mbox{\boldmath $P$}}
\newcommand{\bfQ}{\mbox{\boldmath $Q$}}

\newcommand{\refs}[1]{(\ref{#1})}
\newcommand{\ind}{1\hspace{-1mm}{\rm I}}

\begin{document}

\def\spacingset#1{\renewcommand{\baselinestretch}%
{#1}\small\normalsize} \spacingset{1}


\if0\blind

{
  \title{\bf Modelling of lung cancer survival data for critical illness insurances}
  \author{Joanna D\c{e}bicka \thanks{
    Corresponding author. E-mail: joanna.debicka@ue.wroc.pl.
    This work was partially supported by The National Science Centre Poland under Grant
    2013/09/B/HS4/00490 }\\
    Department of Statistics, Wroclaw University of Economics\\
    and \\
   Beata Zmy\'{s}lona \\
    Department of Statistics, Wroclaw University of Economics}
  \maketitle
} \fi

\if1\blind
{
  \bigskip
  \bigskip
  \bigskip
  \begin{center}
    {\LARGE\bf Modelling of lung cancer survival data for critical illness insurances}
\end{center}
  \medskip
} \fi

\bigskip
\begin{abstract}
We derive a general multiple state model for critical illness
insurances. In contrast to the classical model,
we take into account that the probability of death
for a dread disease sufferer may depend on the duration of the
disease, and the payment of benefits associated with a severe
disease depends not only on the diagnosis but also on the
disease stage.
We apply the introduced model to the analysis of a
critical illness insurance against the risk of lung cancer.
Based on the real data for the Lower Silesian Voivodship in Poland,
we estimate the transition matrix, related to the discrete-time Markov model.
The obtained probabilistic structure of the model can be directly used
to cost not only critical illness insurances and
life insurances with accelerated death benefits  option, but also
to viatical settlement contracts.

\end{abstract}

\noindent%


\spacingset{1.15} 

\section{Introduction}\label{sec:intro}

Modelling of critical illness survival data being primary developed
in the context of, e.g. health insurance contracts, also plays an important role in the currently analysed problems related to viatical market and life insurances with an accelerated death benefits option.

Critical illness insurances (CII) called also dread disease
insurances (DDI) are typical examples of limited-coverage health
insurance products. They provide the policyholder with a benefit
in case of dread disease, which are included in a set of illnesses
specified by the policy conditions such as heart attack, cancer or
stroke (see \cite{DG93}, \cite{HP99}, \cite{Pit94}, \cite{Pit14}).
Such insurance policies can be shaped in several different ways
for instance depending on the specific insurance market. The basic
benefit is a lump sum benefit, which is paid on diagnosis of a
specified condition, rather than on disablement. The other type of
benefit consists of a set of fixed-amount benefits (annuity
payments). It is worth noting that CII policy does not meet any
specific needs and doesn't protect the policyholder against such
financial loss as of earnings, reimbursement of medical or other
expenses incurred. The insured can use the obtained benefits for
any purpose.

Nowadays, due to the growth of the secondary market,
increase of interest
in products providing an acceleration
benefit in a situation related to terminal illness has been observed.
An insured person who has a life
insurance would like to use it when he has financial
problems connected with health. In such situation, the easiest way
of receiving financial compensation from life insurance is to
withdraw from the insurance contract. Then the insurer is
obliged to pay the surrender value. The insured can also sell
his/her policy on the secondary market
for an amount that is greater than the surrender
value (and less than the death benefit).
Then the viatical company takes over the payment of the insurance
premiums, and in case of death of the insured, it gets the death
benefits. Such agreements, called {\it viatical settlements}
(see e.g. \cite{Bhu09}, \cite{Gat10}, \cite{Nee03}),
are offered to people who have developed
a terminal disease. The other
possibility for the insured to receive prior financial
gratification is to buy a life insurance
with an ADBs option that allows the insured to obtain the
death benefit when he is still alive. Sometimes insurance
companies allow the insured to re-buy the option of an accelerated
payment of death benefits after the diagnosis of the disease. This
flexibility is stimulated by the strongly growing viatical market
for life insurance.

A statistical model for survival analysis is equivalent to a
two-state Markov process with one direct transition from a
transient {\it alive} state to an absorbing {\it death} state.
This model is insufficient in framework of study and analysing the
detailed life history data which occur frequently in practice, as
for example in CII. In the literature, depending on the analysed
problem, there have been observed two basic approaches of
designing a suitable model.
On the one hand, the {\it alive} state
can be split into two or more transient states which, in
applications, typically correspond to occurrences of various
medical complications, like for example in case of an acute myocardial infarction (\cite{HM85}) or
insulin-dependent diabetes (\cite{And88}). On the other hand, the
{\it death} state can be split into two or more absorbing states,
which in applications typically correspond to analysing causes of
death and the competing risks survival analysis. The basic model
for CII (e.f. \cite{HP99}, \cite{Pit94}, \cite{Pit14}) combines
both approaches, but does not include the specific terms of
contracts offered by insurers and is not suitable for
costing viatical products and life insurances with ADBs option.

The aim of this contribution is two-fold. In the first part we
present a general multiple state model for critical illness
insurances, which takes into account that a probability of death
for a dread disease sufferer may depend on the duration of the
disease and the payment of benefits associated with a severe
disease are related to a diagnosis and the disease stage.
Due to the nature of the analyzed products,
we propose to split the death state in a different way than in the basic model for CII.

In order to cost insurance and viatical contract, the probabilistic structure
is necessary. Then, in the second part, we focus on the modelling
of the probabilistic structure of the proposed multiple state
model for \emph{products} associated with the risk of lung cancer. For
determining the transition matrix related to the discrete-time
Markov model, we use the methodology developed for the
construction of multi-state life tables. In particular, the
dependence of the survival probability on a current age is
modelled by the logistic regression model for ordered categorical
variable and Poisson regression model with identity link
function. Probability of metastasis diagnosis is expressed using
logit model. The numerical results are based on the actual data for
the Lower Silesian Voivodship in Poland.
Thus we derive the transition matrix
for the proposed model,
which can be used to cost CII contracts, life insurances with ADBs option and viatical settlements.

The paper is organized as follows. In Section \ref{sec:mm}, after
a brief description of the classical multiple state model for the
CII, we propose a more general multiple state model for such
insurances. Section \ref{sec:prob.structure} is a study of the
describtion of the probabilistic structure of the introduced
model. Then in Section \ref{sec:lung.cancer} we focus on the
insurance against the risk of lung cancer, which is an example of
the CII. Based on the actual data for the Lower Silesian
Voivodship in Poland, we estimate transition probabilities for CII
associated with the lung cancer in Section \ref{sec:E.p.t}.
In Section \ref{sec:conclusion} we point out
obtained results to possible applications in practice.

\section{An actuarial model for critical illness insurance}\label{sec:mm}

Following Haberman \& Pitacco \cite{HP99} with a given insurance
contract we assign a {\it multiple state model}. That is, at any
time the insured risk is in one of a finite number of states
labelled by $1,2,...,N$ or simply by letters. Let ${\mathcal{S}}$
be the {\it state space}. Each state corresponds to an event which
determines the cash flows (premiums and benefits). Additionally,
by ${\mathcal{T}}$ we denote the {\it set of direct transitions}
between states of the state space.  Thus ${\mathcal{T}}$ is a
subset of the set of pairs $\left( {i,j} \right)$, i.e.,
${\mathcal{T}} \subseteq \{  \left( {i,j} \right)\mid i \neq j;
i,j \in {\mathcal{S}} \}$. Note that the pair
$({\mathcal{S}},{\mathcal{T}})$ is called a {\it multiple state
model}, and describes all possible insured risk events as far as
their evolution is concerned (usually up to the end of insurance).
In this paper we consider an insurance contract issued at time $0$
(defined as the time of issue of the insurance contract) and
according to a plan terminating at a later time $n$ ($n$ is the
term of policy). Moreover, let $x$ be the age at the policy issue.

\begin{figure}
\hspace{4cm}
\includegraphics[width=5cm]
{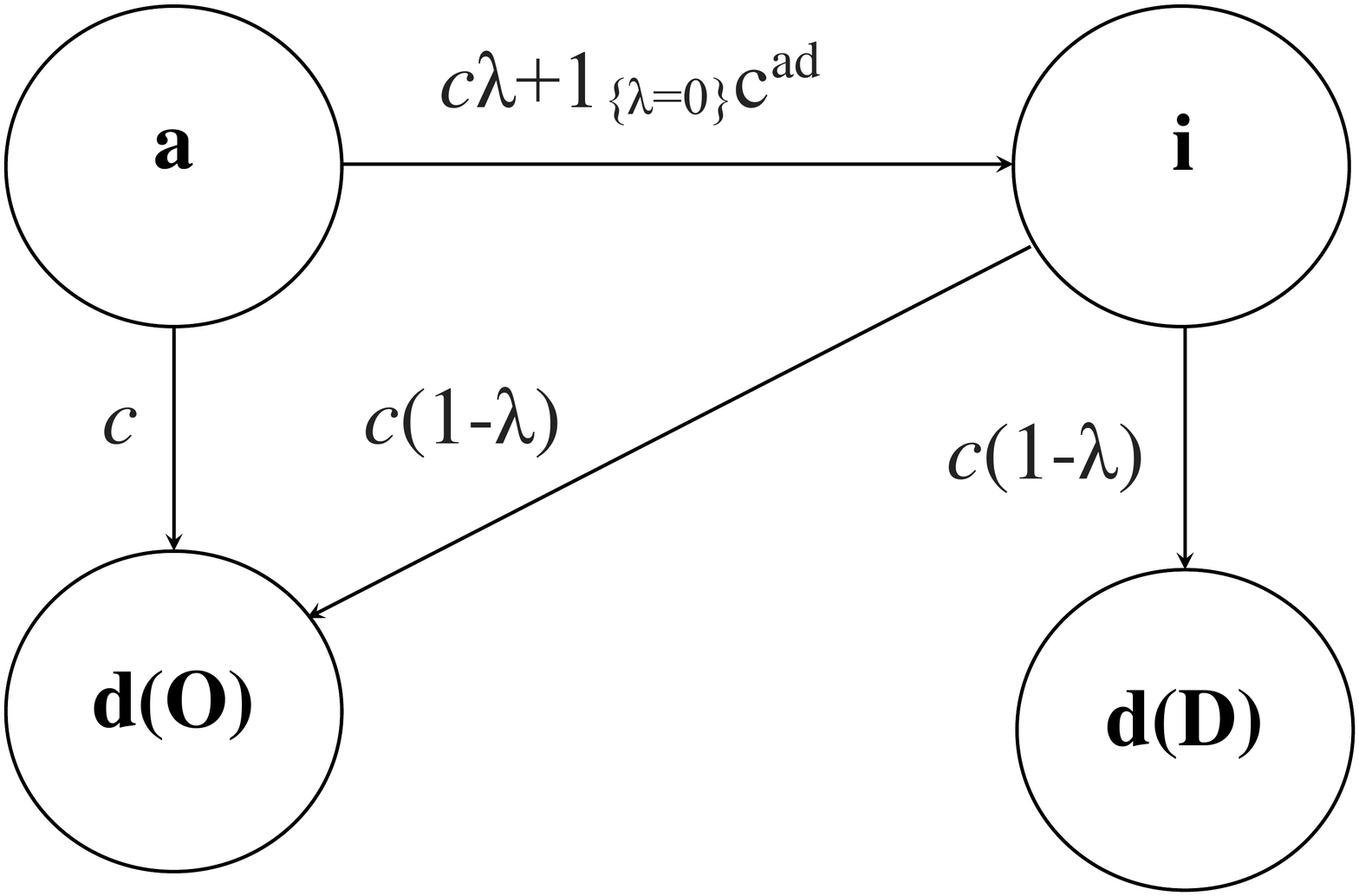}
\caption{\label{Fig.1}A multiple state model for dread disease
insurance with benefits.}
\end{figure}

The most basic multiple state model for CII,
analysed in \cite{DG93}, has the following form
\begin{eqnarray}
(\mathcal{S},\mathcal{T})=(\{a,i,d\},\{(a,i),(a,d),(i,d)\}), \nonumber
\end{eqnarray}
where $a$ means that the insured is active or healthy, $i$  indicates that
the insured person is ill and suffers from a dread disease and $d$
is related to the death of the insured.
A more advanced model was investigated in \cite{HP99}, \cite{Pit94}, \cite{Pit14},
where instead of a single state $d$, it distinguishes
between death being due to dread disease $d(D)$ and other causes
$d(O)$
\begin{eqnarray}\label{DD-MSM}
(\mathcal{S},\mathcal{T})=(\{a,i,d(D),d(O)\},\{(a,i),(a,d(O)),(a,d(D))(i,d(O)),(i,d(D))\}.
\end{eqnarray}
A multiple state model for such a critical
illness cover is presented in Figure~\ref{Fig.1}. Next to the arcs,
benefits related to the transition between states are marked,
where $c$ is a given lump sum ({\it death benefit}), $c^{ad}$ is
an additional lump sum ({\it disease benefit}) and $\lambda$ is
the so called {\it acceleration parameter} $(0 \leq \lambda \leq
1)$. The amount $c\lambda+\ind_{\{\lambda=0\}} c^{ad}$ is payable
after the dread disease diagnosis, while the remaining amount
$c(1-\lambda)$ is payable after death, if the two random events
occur within the policy term $n$.

Note that the multiple state model \refs{DD-MSM} covers all
forms of critical illness insurances, namely
\begin{description}
\item[-] if $\lambda=0$, then the model describes a rider benefit as an additional benefit,
\item[-] if $0< \lambda <1$, then the model describes a rider benefit as an acceleration of part of
the basic life cover,
\item[-] if $\lambda=1$, then the model describes a stand-alone cover.
\end{description}
If $\lambda=1$, the state $i$  is absorbing, because the
whole insurance cover ceases immediately after dread disease
diagnosis and the payment of the sum assured (in this model
direct transitions $(i,d(O))$ and $(i,d(D))$ are not present).
If $0 \leq \lambda <1$, state $i$ is irreversible.

Costing of any insurance products is always connected with the
probability structure of the model. In case of dread disease
cover, data such as incidence rates of dread disease is required.
A frequent problem is that the needed data is not available or only in a limited
form. Therefore, one has to make assumptions, which have impact on
the actuarial values such as premiums and reserves. For the CII
designed in Figure~\ref{Fig.1}, the outline of possible
assumptions and methods for calculating premiums rates for a critical illness
cover is presented in \cite{DG93}.  One of the objectives is to
assume that the probability of death of a sick person does not
depend on the duration of the disease. Moreover, the dread disease insurances
are products, which are very sensitive to the development of
medicine. Not all dread diseases are as mortal as some years ago
and yet this type of insurances are of long-term type. Thus
insurers introduce very strict conditions for the right to receive
the benefit associated with a severe disease. Beside a diagnosis,
the disease stage is important. This implies that the model
presented in Figure~\ref{Fig.1} is insufficient.

In this paper we propose a multiple state model for critical illness
insurance, which takes into account that the probability of death
for a dread disease sufferer may depend on the duration of the
disease and the payment of benefits associated with a severe
disease depend on a diagnosis and the disease stage.

\begin{figure}[h!bt]
\hspace{4cm}
\includegraphics[width=8cm
]{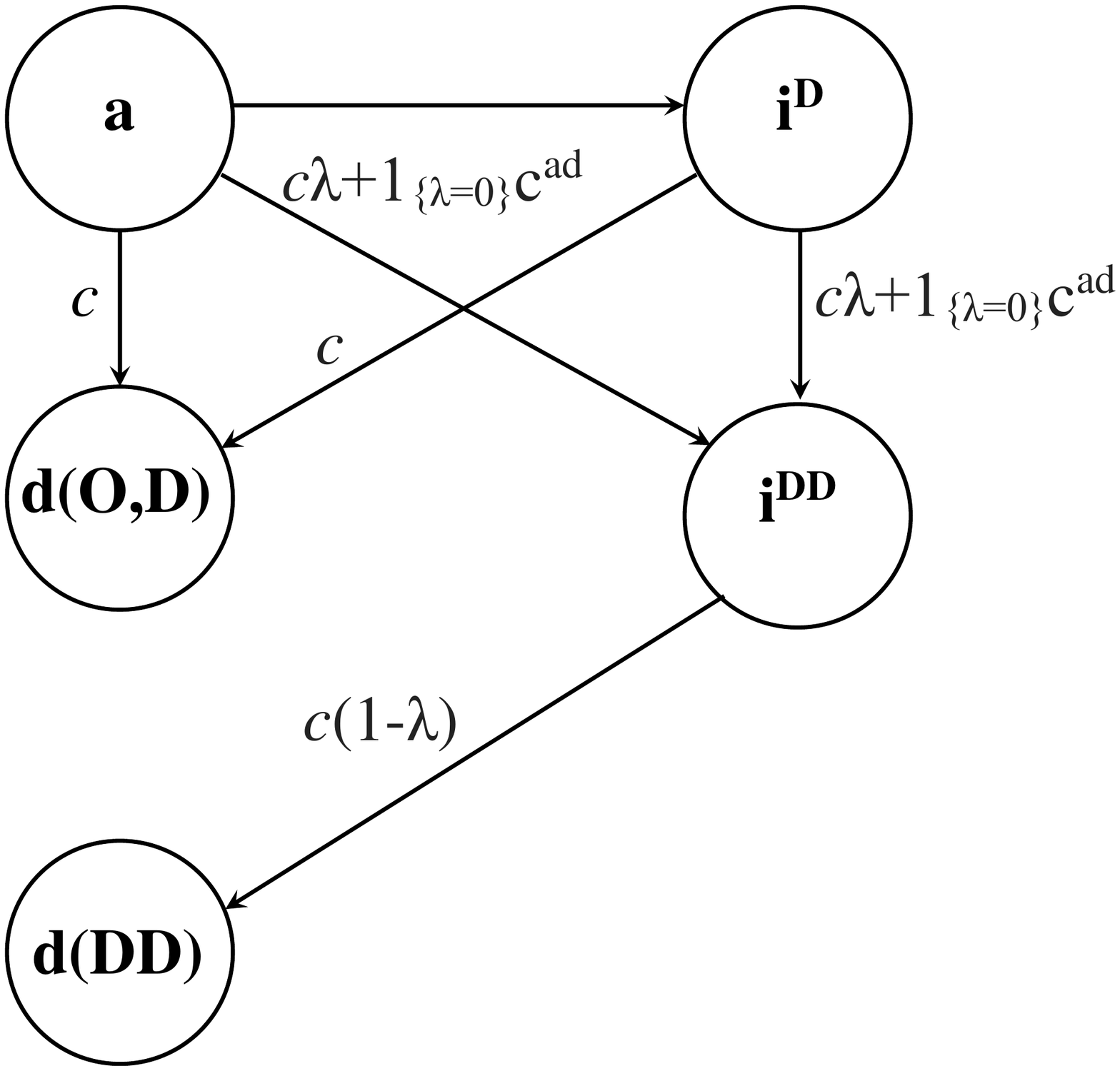}
\caption{\label{Fig.2} A multiple state model for CII with
benefits suited to the real conditions of a contract.}
\end{figure}

Let $e_s$ be an expected future lifetime of $s$-years-old person.
It is important to know that critical illness benefits are paid
not only on the diagnosis but also on the expected future lifetime
of a sick person. Typically the benefit is paid to a sick person
whose expected future lifetime is no longer than four years or, in
some cases, two years.  These differences come from medical
circumstances. On the one hand, the term {\it terminally ill} in
the context of health care refers to a person who is suffering
from a serious illness and whose life is not expected to go beyond
24 months at the maximum. On the other hand, the HIV+ patients
with more than approximately 4,5 years of life expectancy are
treated as patients in relatively good health. Therefore, in order
to accommodate such a condition in the model of CII, state $i$ has
to be divided into two states:
\begin{description}
\item[$i^D$ -] the insured person is ill and his
expected future lifetime is at least 4 years ($e_s\geq4$). In this
stage the remission of the disease is still possible, although
return to health state is impossible.
\item[$i^{DD}$ -] the insured
person is ill and his expected future lifetime is less than
4~years ($e_s<4$). In this stage the remission of the disease is
very unlikely.
\end{description}
After such a division a multiple
state model for CII covers has a form presented in
Figure~\ref{Fig.2} where
\begin{description}
\item[$d(O,D)$ -] the death of the insured person who is ill and his
expected future lifetime is at least 4 years ($e_s\geq4$) or due to other cases.
\item[$d(DD)$ -] the death of the insured
person who is ill and his expected future lifetime is less than
4~years ($e_s<4$).
\end{description}

Note that in the model presented in Figure~\ref{Fig.2} we describe
death states in a different way than in Figure~\ref{Fig.1}.
The main reason is that in described CII, values of benefits are connected
with the insured's health situation just before his death, not with the cause of death.

Moreover, the direct transition $(a,d(DD))$ is omitted.
In particular, this means that all dread
disease deaths of the insured with expected future lifetime less than
4~years  are represented by the state $d(DD)$. It is a
situation analogous to the one given in Figure~\ref{Fig.1} where deaths
due to dread disease are represented by the pair of transitions
$(a, i)$ and $(i, d(D))$, the direct transition $(a, d (D))$ is
not possible.

To avoid a situation of 'overpayment' (that
could take place when death occurs within a very short period
after disease inception to the terminal phase of the dread
disease i.e. $i^{DD}$), the single cash payment $c^{ad}$ is replaced with a
series of payments  $b$ (the annuity).
So, it seems reasonable to assume that the probability of death for a
dread disease sufferer depends on the duration of the disease. To
consider the influence of illness duration on the mortality
probabilities, we split state $i^{DD}$ into four states
$i^{DD(h)}$ $(h=1,2,3,4)$, where $i^{DD(h)}$ means that the
insured is terminally sick and his expected lifetime is less
than $4-(h-1)$ years (compere \cite{Ams68}, \cite{Gre93},
\cite{HP99}, \cite{JM06}, \cite{JM07}). Note that state
$i^{DD(h)}$ is a reflex state (that is strictly transitional and
after one unit of time, the insured risk leaves this state).
\vspace{0.5cm}

\begin{figure}[bht]
\hspace{0.01cm}
\includegraphics[width=15cm]{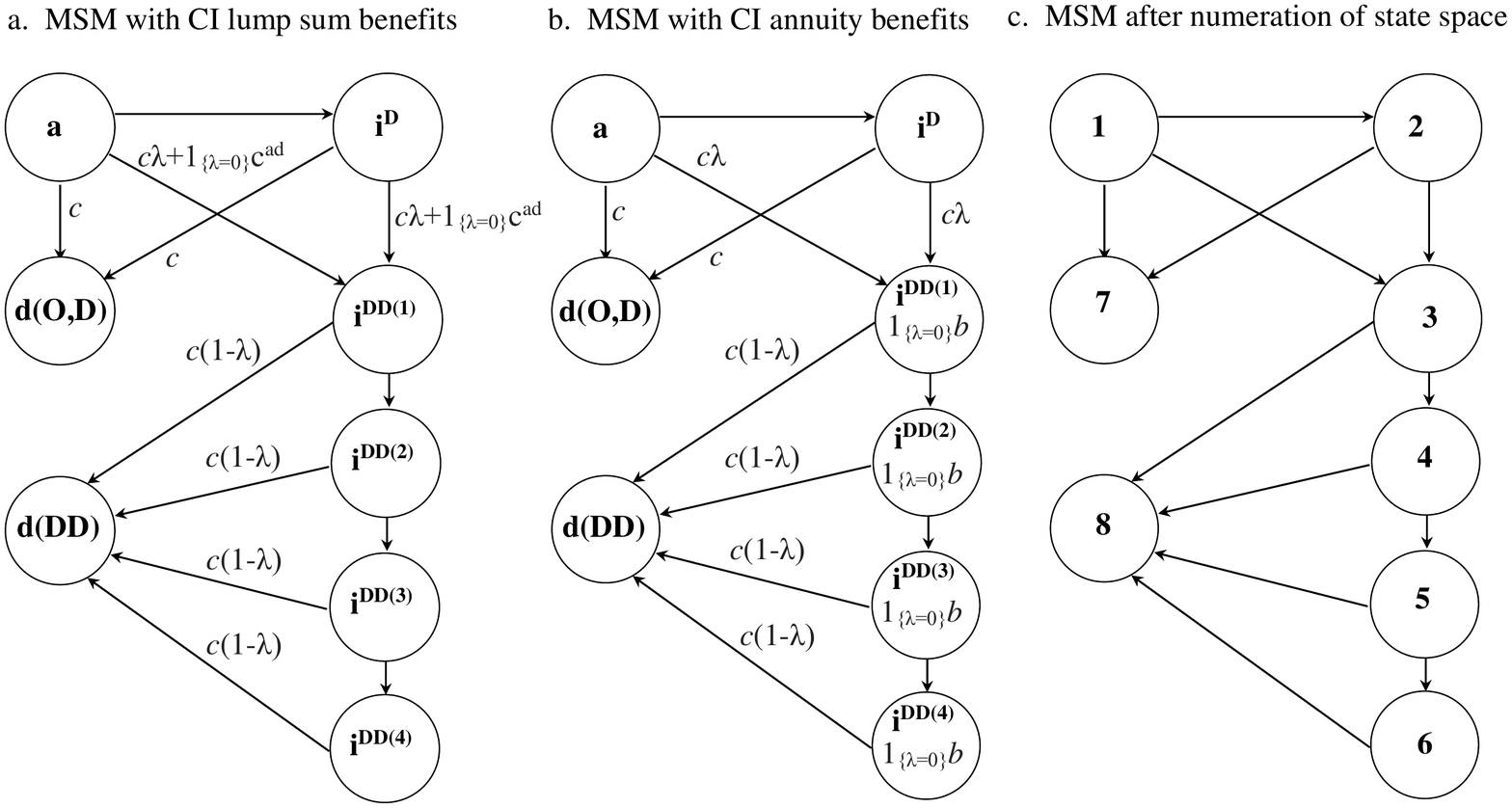}
\caption{\label{Fig.8} A general multiple state model for CII.}
\end{figure}
Finally, we arrive at a general multiple state model for critical illness insurances,
which covers both disease lump sum benefits (Figure~\ref{Fig.8}a) and
disease annuity benefits (Figure~\ref{Fig.8}b). In particular,
the model presented in Figure~\ref{Fig.8}b may be applied to CII contract
with  increasing ($b_3<b_4<b_5<b_6$) or decreasing ($b_3>b_4>b_5>b_6$) annuity benefits,
where $b_j$ is an annuity rate realized at state $j$ for $j=3,4,5,6$. The model with the
state space after numeration can be seen in Figure~\ref{Fig.8}c.

\section{Probabilistic structure of the model}\label{sec:prob.structure}

\subsection{Decrements and assumptions}\label{subsec:decrements.assumprions}

We focus on a discrete-time model. Let $X(x,t)$ denote the state
of an individual (the policy) at time t ($t \in \textrm{T}=
\{0,1,2,\dots ,n\}$), where $x$ is an {\it age at entry}. Hence
the evolution of the insured risk is given by a discrete-time
stochastic process $\{ {X(x,t); t\in \textrm{T}} \}$, with values
in the finite set $\mathcal{S}=\{ 1, 2,...,N\}$. The individual's
presence in a given state or movements between states
may have some financial impact like payments of premiums or benefits.
In order to valuate the
insurance contract during $n$-years insurance period, the
knowledge of probabilities of realizing particular cash flows is
necessary.

To describe the probabilistic structure of $\{X(x,t)\}$, for any
moment $k \in \{0,1,2,...,n \}$, we introduce vector
$$ \bfP^{[x]}(k) =(\Prob^{[x]}_{1}(k), \Prob^{[x]}_{2}(k), \Prob^{[x]}_{3}(k), \ldots
,\Prob^{[x]}_{N}(k))^T \in \RL^{N}, $$ where
$\Prob^{[x]}_{j}(k)=\Prob(X(x,k)=j)$. Note that
$\bfP^{[x]}(0)=\bfP(0) \in \RL^{N}$ is a vector of the initial
distribution (usually it is assumed that state $1$ is an initial
state, that is $ \bfP(0) =(1,0,0, \ldots ,0)^T \in \RL^{N}$ for
each $x$).

Under the assumption that $\{X(x,t)\}$ is a nonhomogeneous Markov chain,
to get the sequence of matrices
$\{\bfP^{[x]}(k)\}_{k=0}^{n}$, it is enough to know $\bfP(0)$ and
the sequence of matrices $ \bfQ^{[x]}(0),
\bfQ^{[x]}(1),\bfQ^{[x]}(2), \ldots, \bfQ^{[x]}(n-1)$, where
$\bfQ^{[x]}(k)=\left(q^{[x]}_{ij}(k)\right)_{i,j=1}^{N}$ and
$q^{[x]}_{ij}(k)=\Prob (X(x,k+1)=j | X(x,k)=i)$ is a transition
probability.

A transition matrix of $\{X(x,t)\}$ for CII model given in
Figure~\ref{Fig.8} has the following form
\begin{eqnarray}\label{macierzQ}
\bfQ^{[x]}(k)= \left(
 \begin{array}{cccccccc}
         q^{[x]}_{11}(k) & q^{[x]}_{12}(k)  & q^{[x]}_{13}(k)  &                0 &                0 &                0 &  q^{[x]}_{17}(k) &                 0  \\
                       0 & q^{[x]}_{22}(k)  & q^{[x]}_{23}(k)  &                0 &                0 &                0 &  q^{[x]}_{27}(k) &                 0  \\
                       0 &                0 &                0 &  q^{[x]}_{34}(k) &                0 &                0 &                0 &    q^{[x]}_{38}(k)  \\
                       0 &                0 &                0 &                0 &  q^{[x]}_{45}(k) &                0 &                0 &    q^{[x]}_{48}(k)  \\
                       0 &                0 &                0 &                0 &                0 &  q^{[x]}_{56}(k) &                0 &    q^{[x]}_{58}(k)  \\
                       0 &                0 &                0 &                0 &                0 &                0 &                1 &                 0  \\
                       0 &                0 &                0 &                0 &                0 &                0 &                1 &                 0  \\
                       0 &                0 &                0 &                0 &                0 &                0 &                0 &                 1
 \end{array}
 \right)
\end{eqnarray}

In the next section we provide formulas for $q^{[x]}_{ij}(k)$.

\subsection{Multiple increment-decrement tables}\label{subsec:i.d.tables}

Transition probabilities can be determined using a {\it multiple
increment-decrement table} (or {\it multiple state life table}).
The number of functions of such a table is closely linked to
multiple state model $(\mathcal{S}, \mathcal{T})$ (cf.
\cite{Deb12}, \cite{Piec}, \cite{Hab83a}, \cite{Hab83b},
\cite{Jor82}, \cite{Mat97}). The simplest multiple
increment-decrement table, which refers to an $x$ year old person
is a life table $\{l_{[x]+k}\}_{k \geq 0}$, where $l_{[x]+k}$ is a
number of those alive at the beginning of time interval
$[x+k,x+k+1)$. Then $d_{[x]+k}=l_{[x]+k}-l_{[x]+k+1}$ is the
number of deaths during the time interval $[x+k,x+k+1)$. In
general, one can assign for a multiple state model $(\mathcal{S},
\mathcal{T})$ and an $x$ year old person the multiple
increment-decrement table, which consists of functions described
for each transient state $i \in \mathcal{S}$:
\begin{description}
\item[$l^{i}_{[x]+k}$] - denotes the number of lives in state $i$ at age $x + k$,
\item[$d^{ij}_{[x]+k}$] - the number of lives at
age $x+k$, which during period $[x+k, x+k+1)$ left the state $i$
and transit to state $j$.
\end{description}
The following recurrence relation holds
\begin{eqnarray}\label{l-d:WTTZ}
l^{i}_{[x]+k+1}=l^{i}_{[x]+k} - \sum_{j:(i,j) \in \mathcal{T}}
d^{ij}_{[x]+k} + \sum_{j:(j,i) \in \mathcal{T}} d^{ji}_{[x]+k},
\end{eqnarray}
where $\sum_{j:(i,j) \in \mathcal{T}} d^{ij}_{[x]+k}$ is a number
of lives, which left the state $i$ in time interval $(x+k,x+k+1]$,
and $\sum_{j:(j,i) \in \mathcal{T}}d^{ji}_{[x]+k}$ is a number of
lives, which entered state $i$ in time interval $(x+k,x+k+1]$.

The multiple increment-decrement table
\begin{eqnarray}
\left\{l_{[x]+k}^{1},l_{[x]+k}^{2},l_{[x]+k}^{3},l_{[x]+k}^{4},l_{[x]+k}^{5},l_{[x]+k}^{6},
d_{[x]+k}^{12}, d_{[x]+k}^{13}, d_{[x]+k}^{17}, d_{[x]+k}^{23},
d_{[x]+k}^{27}, \right. \nonumber \\
\left. d_{[x]+k}^{34}, d_{[x]+k}^{38}, d_{[x]+k}^{45},
d_{[x]+k}^{48}, d_{[x]+k}^{56}, d_{[x]+k}^{58} \right\}_{k \geq 0}
\label{mi-dlt}.
\end{eqnarray}
refers to an $x$ year old person for
$({\mathcal{S}},{\mathcal{T}})$ given in Figure~\ref{Fig.8}.

The following relations holds between elements of $\bfQ^{[x]}(k)$ and
functions of multiple increment-decrement table for
$({\mathcal{S}},{\mathcal{T}})$:
\begin{description}
\item[--] if $i$ is absorbing, then
\begin{eqnarray}
q^{[x]}_{ij}(k)= \left\{
\begin{array}{rll}\label{q_absorbing}
1 &{\rm for}& j=i \\
0 &{\rm for}& j \neq i
\end{array}
\right. ,
\end{eqnarray}
\item[--] if $i$ is transient, then
\begin{eqnarray}
q^{[x]}_{ij}(k)= \left\{
\begin{array}{lll}\label{q_transient}
\frac{l^{i}_{[x]+k+1}- \sum_{j:(i,j) \in \mathcal{T}} d^{ij}_{[x]+k}}{l^{i}_{[x]+k}} &{\rm for}& j=i  \\
\frac{d^{ij}_{[x]+k}}{l^{i}_{[x]+k}} &{\rm for}& (i,j) \in {\mathcal{T}}  \\
0 &{\rm for}& (i,j) \notin {\mathcal{T}}
\end{array}.
\right.
\end{eqnarray}
\end{description} 

The preparation of multiple increment-decrement tables for each
age $x$ is cumbersome and not always needed. In further analysis
we suppose that the distribution of $\{X(x,t)\}$ can be expressed
by the distribution of process $\{X(0,s)\}$ with regard
to a $0$ year old person. This assumption is known as the {\it
hypothesis of aggregation} (HA), which can be equivalently
formulated in several ways.
Observe that, according to the model presented in Figure~\ref{Fig.8}c,
the {\it alive} state has been split into six transient
states (i.e. states $1,2,...6$) and the {\it death} state has been
split into two absorbing states corresponding
to health situation of the insured just before his death (i.e. states $7$ and $8$).
Then, HA for the considered model
is equivalent to the condition
\begin{eqnarray}\label{HA}
\bigwedge_{j \in \{1,2,...6\}}
\Prob (X(x,k)=j)&=& \Prob (X(0,x+k)=j \mid X(0,x) \in \{1,2,3,...,6\})
\end{eqnarray}
for $x$ and $k$ that $\Prob (X(x,k)=j)>0$.

Based on \refs{HA}, it can be shown that
\begin{eqnarray}
q^{[x]}_{ij}(k)   &=& \Prob (X(x,k)=j \mid X(x,k-1)=i)  \nonumber \\
            &=& \Prob (X(0,x+k)=j \mid X(0,x+k-1)=i) \nonumber \\
            &=& q^{[0]}_{ij}(x+k) \nonumber
\end{eqnarray}
so we obtain $\left\{ \bfQ^{[x]}(k)\right\}^{n-1}_{k=0}=\left\{ \bfQ^{[0]}(x+k)\right\}^{n-1}_{k=0}$.

In order to simplify the notation, let $\{X(s)\}:=\{X(0,s)\}$
and for given $x$ we have $\bfQ(k):=\bfQ^{[0]}(x+k)$ with
\begin{eqnarray}
q_{ij}(k)=\Prob (X(x+k+1)=j \mid X(x+k)=i). \nonumber
\end{eqnarray}
Probabilities $q_{ij}(s)$, $s\geq0$ can be calculated in the same way as
in \refs{q_absorbing} and \refs{q_transient} but using the
multiple increment-decrement table
\begin{eqnarray}
\left\{l_{s}^{1},l_{s}^{2},l_{s}^{3},l_{s}^{4},l_{s}^{5},l_{s}^{6},
d_{s}^{12}, d_{s}^{13}, d_{s}^{17}, d_{s}^{23}, d_{s}^{27},
d_{s}^{34}, d_{s}^{38}, d_{s}^{45}, d_{s}^{48}, d_{s}^{56},
d_{s}^{58} \right\}_{s \geq 0} \label{HAs.mi-dl}.
\end{eqnarray}
Unfortunately, an appropriate set of data allowing to create the
multiple increment-decrement table \refs{HAs.mi-dl} is not always
available. If so, estimation of $\bfQ(k)$ is needed. In
Section~\ref{sec:lung.cancer} and Section~\ref{sec:E.p.t}
we focus on this problem in case of lung cancer disease.

\section{Lung cancer as an example of dread disease}\label{sec:lung.cancer}

After cardiovascular diseases, malignant tumors pose the second
cause of death in developed countries. In particular, lung cancer
belongs to the group of tumors characterized by the highest
morbidity and mortality rates. It is the most frequent in
population of men and the second frequent in population of women
after breast cancer. Additionally, lung cancer is  so-called tumour
with unfavourable prognosis. For example in Poland, by analysing
epidemiological data it can be concluded that only about $16 \%$
of women and $11 \%$ of men outlive five years after the diagnosis
\cite{WD14}. Because of the high prevalence and mortality rates,
the relatively short survival time after the diagnosis, lung
cancer is a perfect example of the deadly disease, which should be
covered by critical illness insurances.

Epidemiological data confirms the existence of significant
differences between the incidence of lung cancer in men and women
populations. The morbidity rate is several times higher in men
population. In many European countries, in the second part of the
eighties of the XX century, the tendency of stabilizing the
incidence rate is observed among men. A different situation occurs
in case of women. The number of cancer cases continues to grow,
which is undoubtedly caused by cultural changes, such as an
increase in the number of smokers among women in the post-war
generation. Due to the growing number of smokers among women, we
should expect a further increase in the number of cancer cases in
the population of women.

The incidence rate depends also on age. Lung cancer occurs very
rarely among  patients up to forty years of age. The incidence
begins to increase after the age of fifty. The peak incidence
occurs at the sixth and seventh decades of life. By analysing
geographical data, a significant diversity of incidence and
mortality rates is observed in different regions of Europe. In
Poland, the morbidity and mortality vary significantly among
particular provinces (voivodships). Thus, age, sex and region of
residence should be taken into account in the analysis of the
etiology of lung cancer.

In case of CII for lung cancer, the model (presented in
Figure~\ref{Fig.8}) has six states associated with health
situation of the insured person which mean that the insured:
\begin{description}
\item[$1$] - is alive and not sick with malignant lung tumour,
\item[$2$] - is diagnosed of lung cancer without
metastasis to lymph nodes, brain, bones or so-called distant
metastases, found,
\item[$3$] - is diagnosed of lung cancer and the existence of
distant metastases is observed and his expected lifetime is
less than $3$ years ($e_s < 4$),
\item[$4$] - has a lung cancer with distant metastases and $e_s < 3$,
\item[$5$] - has a lung cancer with distant metastases and $e_s < 2$,
\item[$6$] - has a lung cancer with distant metastases and $e_s < 1$,
\end{description}
Other states are associated with the death of the insured person who, before his death:
\begin{description}
\item[$7$] - was healthy or is diagnosed of lung cancer without
metastasis,
\item[$8$] - had a lung cancer with distant metastases.
\end{description}

\section{Estimation of transition probabilities}\label{sec:E.p.t}

\subsection{Data}
Due to the influence of the residence place on morbidity and
mortality rates of lung cancer an analysis based on data from
Lower Silesia separately for men and women populations was
performed. In order to estimate the transition probabilities three
databases have been used.

First, in the analysis of future life time, the life tables for
2008, separately for population of men and women were used
\cite{TTZ-GUS}.

Secondly, the information about the morbidity and mortality rates
is obtained on the basis of the National Cancer Registry for the
Lower Silesia Region \cite{WD14}. The register is created on the basis
of individual declarations of tumors by hospitals.
Note that, in the year 2008, the
percentage of the declarations submitted in Lower Silesia region
exceeds 95\% and belongs to the best registries in Poland.
Therefore, this database is reliable.

In the analysis, the data set of individual hospitalization from
the Lower Silesia Department of the National Health Fund was used
\cite{NHF}.  The number of patients with
lung cancer was identified using the disease code (C33 and C34
according to the system of codes from ICD-10). Patients were
identified using the coded numbers of the Universal Electronic
System for Registration of the Population (Social Security). Data
for the period from 2006 to 2011 was included in the analysis. The
year 2008, as one of the middle periods, has been established as
the reference year. The choice of the middle period allows to
consider the histories of hospitalization of these patients in the
time horizon from 2006 to 2011.

Populations of men and women are examined separately, due to a
different structure of the morbidity and mortality associated with
lung cancer. A data set concerning histories of hospitalization
because of malignant lung cancer in 2008 in 62 hospitals was used.
In the entire Lower Silesia Voivodeship, 2246 men (at age 20-94)
and 945 women (at age 23-93) were hospitalized.

Since we do not know the exact date of death, we have only
knowledge about a cessation of traditional treatment and the
transfer of a patient to a hospice, the survival time is
determined with an accuracy of a year.

\subsection{Active}

This section concerns probabilities associated with state $1$ of
the CII model i.e. probabilities of the first row of the matrix
\refs{macierzQ}

Due to \refs{q_transient}, under HA, we
obtain
$$q_{11}(k)=\frac{l^{1}_{x+k}-(d^{12}_{x+k}+d^{13}_{x+k}+d^{17}_{x+k})}{l^{1}_{x+k}}, q_{12}(k)=\frac{d^{12}_{x+k}}{l^{1}_{x+k}},q_{13}(k)=\frac{d^{13}_{x+k}}{l^{1}_{x+k}},q_{17}(k)=\frac{d^{17}_{x+k}}{l^{1}_{x+k}}.$$
Note that the probability of developing lung cancer without
detected metastases can be decomposed as follows
\begin{eqnarray} \label{q_12}
q_{12}(k)=\left({\frac{d^{12}_{x+k}+d^{13}_{x+k}}{l_{x+k}}}\right)/\left({\frac{l^{1}_{x+k}}{l_{x+k}}}\right)
         -\left({\frac{d^{13}_{x+k}}{d^{12}_{x+k}+d^{13}_{x+k}}}\right)/\left({\frac{l^{1}_{x+k}}{d^{12}_{x+k}+d^{13}_{x+k}}}\right).
\end{eqnarray}
The expression $(d^{12}_{x+k}+d^{13}_{x+k})/l_{x+k}$ (called the
morbidity rate) denotes the ratio of the number of people who fell
ill to the whole population. It is calculated on the basis of the
crude cancer incidence rate using data from \cite{WD14}. Let
$\zeta_{s}^{(t)}=\frac{\check{l}_{s}^{(t)}}{100000}$ denote the
crude cancer incidence rate for $t$-th year as the number of
cases of illness $\check{l}_{s}^{(t)}$ per 100000 of the studied
population at age $s$, calculated in five-year age groups. Because
$\zeta_{s}^{(t)}$ has significant variability, therefore we used
the average of the crude cancer incidence rates
\begin{eqnarray}
\bar{\zeta}_{s}=\frac{1}{5} \sum_{t=2006}^{2010}\zeta_{s}^{(t)}.
\end{eqnarray}

The estimation of $(d^{13}_{x+k})/(d^{12}_{x+k}+d^{13}_{x+k})$
required to separate a cohort of patients who, in 2008, had lung
cancer diagnosed. Patients are divided into two groups. Patient
with metastases during the first diagnosis belong to the first
group. The second group consists of patient without metastases
during the first diagnosis. Let  $\beta_{s}$ denote the percentage
of $s$ year old patients who fell ill in 2008 with the first
diagnosis showing the existence of distant metastases is
calculated in five-year age groups. It can easily be seen that the
proportion of people without lung cancer and the whole population
is close to one, which allows an assumption that
\begin{eqnarray} \label{ap_1}
{\frac{l^{1}_{s}}{l_{s}}}\approx 1.
\end{eqnarray}
Then the probability \refs{q_12} takes the following form
\begin{eqnarray} \label{est_q_12}
q_{12}(k)=\bar{\zeta}_{x+k}(1-\beta_{x+k}).
\end{eqnarray}
By the same argument
\begin{eqnarray} \label{est_q_13}
q_{13}(k)=\bar{\zeta}_{x+k} \cdot \beta_{x+k}.
\end{eqnarray}

To estimate the probability of death of a healthy person for the
reason other than lung cancer, the crude cancer mortality rate
${\varpi}_{s}$ should be defined as a number of deaths
$\check{d}_{s}$ per $100000$ of the studied population at age $s$
(calculated in five-year age groups) and can be expressed as
follows
\begin{eqnarray}\label{varpi}
{\varpi}_{s}=\frac{d_{s}-d^{17}_{s}}{l_{s}}.
\end{eqnarray}
Hence by \refs{ap_1} and \refs{varpi} we arrive at
\begin{eqnarray} \label{q_17}
q_{17}(s) =\left(\frac{d_{s}}{l_{s}}-{\varpi}_{s}\right)/
\left(\frac{l^{1}_{s}}{l_{s}}\right)=q_{s}-{\varpi}_{s}.
\end{eqnarray}
Following the same procedure as for $\zeta_{s}^{(t)}$, let
\begin{eqnarray}
\bar{\varpi}_{s}=\frac{1}{5} \sum_{t=1}^{5}\varpi_{s}^{(t)}
\end{eqnarray}
where $\varpi_{s}^{(t)}=\frac{\check{d}_{s}^{(t)}}{100000}$ is the
crude rates for $t$-th year obtained on basis of
data from \cite{WD14}. We
finally obtain the following formula
\begin{eqnarray} \label{est_q_17}
q_{17}(k)=q_{x+k}-\bar{\varpi}_{x+k},
\end{eqnarray}
 where $q_{x+k}={d_{x+k}}/{l_{x+k}}$ is the probability of dying during
the time interval $[x+k,x+k+1)$ calculated on the base of
a life table.

Noting that the sum of the transition probabilities from a given
state is equal to one, then using \refs{est_q_12}, \refs{est_q_13}
and \refs{est_q_17} we obtain
\begin{eqnarray} \label{est_q_11}
q_{11}(k)=1- (q_{x+k}-\bar{\varpi}_{x+k})-\bar{\zeta}_{x+k}.
\end{eqnarray}

In the second part of this section, we present methods for
estimation of $\bar{\zeta}_{s}$, ${\beta}_{s}$ and
$\bar{\varpi}_{s} $ for $s=20, 21,...100$.

The average crude rates of morbidity and mortality from lung
cancer estimated on the basis of raw indicators from the years
2006-2010 are shown in Table~\ref{Tab.1}. The rates were estimated
on the basis of reports from the National Cancer Registry
\cite{WD14}.
\begin{table}[h!bt]
 \begin{center}\label{Tab.1}
  \caption{Crude average mortality and morbidity rates
(based on data from the years 2006-2010) per 100 000
inhabitants of Lower Silesia}
  \vspace{0,1cm}
{\small
\begin{tabular}{|c|c|c|c|c|}
\hline Age ($s$) & $\bar{\zeta}_{s}$ for men & $\bar{\varpi}_{s} $
for men & $\bar{\zeta}_{s} $ for women & $\bar{\varpi}_{s} $ for
women
\\ \hline 20-25 & 0.00000368 &0.00000180 & 0.00000500 & 0.00000050
\\ \hline 25-30 & 0.00000628 &0.00000180 & 0.00000250 & 0.00000110
\\ \hline 30-35 & 0.00001542 &0.00000654 & 0.00001638 & 0.00000708
\\ \hline 35-40 & 0.00004006 &0.00002646 & 0.00002814 & 0.00001744
\\ \hline 40-45 & 0.00013030 &0.00010948 & 0.00008798 & 0.00007262
\\ \hline 45-50 & 0.00037192 &0.00035406 & 0.00020522 & 0.00016596
\\ \hline 50-55 & 0.00102374 &0.00090890 & 0.00045766 & 0.00039284
\\ \hline 55-60 & 0.00182494 &0.00170202 & 0.00088310 & 0.00071290
\\ \hline 60-65 & 0.00302086 &0.00297606 & 0.00109398 & 0.00097314
\\ \hline 65-70 & 0.00430094 &0.00432944 & 0.00112206 & 0.00107720
\\ \hline 70-75 & 0.00528114 &0.00584450 & 0.00114504 & 0.00119580
\\ \hline 75-80 & 0.00565362 &0.00671536 & 0.00118616 & 0.00129806
\\ \hline 80-85 & 0.00458130 &0.00604818 & 0.00107820 & 0.00132638
\\ \hline above 85 & 0.00369674 &0.00477544 & 0.00112764 &
0.00136514\\ \hline
\end{tabular}}
 \end{center}
\end{table}

In order to estimate the percentage of people who fell ill in 2008
and were diagnosed with metastatic disease, a cohort of patients
with lung cancer who, during 2008, fell ill with lung cancer has
been separated. In that year, 1353 men and 605 women were
diagnosed with lung cancer in the region of Lower Silesia.
Patients received one of the two diagnoses. The first
option was a recognition of metastases to lymph nodes in the chest
and so called distant metastases. The analysis included an
additional period of four weeks after making the first diagnosis.
This period, treated as the additional time which is required to
obtain the results of diagnostic tests, is taken into account in
the model. If, during this period, the existence of metastasis was
confirmed, the patient was classified to the same group as
patients who received a diagnosis of metastatic disease during the
first visit. From a formal point of view, the patient moved at
once from the first state to the third.

The diagnosis stating only the tumour incidence in the lungs but
without metastasis was identified as the second possible type of
diagnosis. In the considered model it is assumed that a patient
with a diagnosis of the absence of metastasis goes into the second
state.

Due to the fact that the incidence and mortality rates of lung
cancer from the National Cancer Registry are presented in
five-year age groups, the percentage of people diagnosed with
metastases was also estimated in such groups. The patients in the
age group of 20 to 40 years pose an exception. In this age group
lung cancer occurs extremely rarely, therefore a twenty-year age
limit was used for the estimation. On the basis of data set from
\cite{NHF}, the percentage of patients with diagnosable
metastases during the first visit is shown in Table~\ref{Tab.2}.
\begin{table}[h!bt]
 \begin{center}
  \caption{The percentage of patients with
metastases diagnosed during the first visit or within a four-week
period.}
{\small
\begin{tabular}{|l|c|c|}  \hline \label{Tab.2}
Age ($s$) & $\beta_{s}$ for men & $\beta_{s}$ for women \\ \hline
20-40 & 0.23077 & 0.28571 \\ \hline 40-45 & 0.31818 & 0.50000 \\
\hline
45-50 & 0.54412 & 0.41463 \\ \hline 50-55 & 0.54922 & 0.51282 \\
\hline 55-60 & 0.45963 & 0.53988 \\ \hline 60-65 & 0.50166 &
0.45113 \\ \hline 65-70 & 0.49359 & 0.40000 \\ \hline 70-75 &
0.48246 & 0.25472 \\ \hline 75-80 & 0.36777 & 0.36111 \\ \hline
80-85 & 0.34177 & 0.28125 \\ \hline above 85 & 0.33333 & 0.33333\\
\hline
\end{tabular}}
 \end{center}
\end{table}

\begin{figure}[bht]
\hspace{2cm}
\includegraphics[width=11cm]{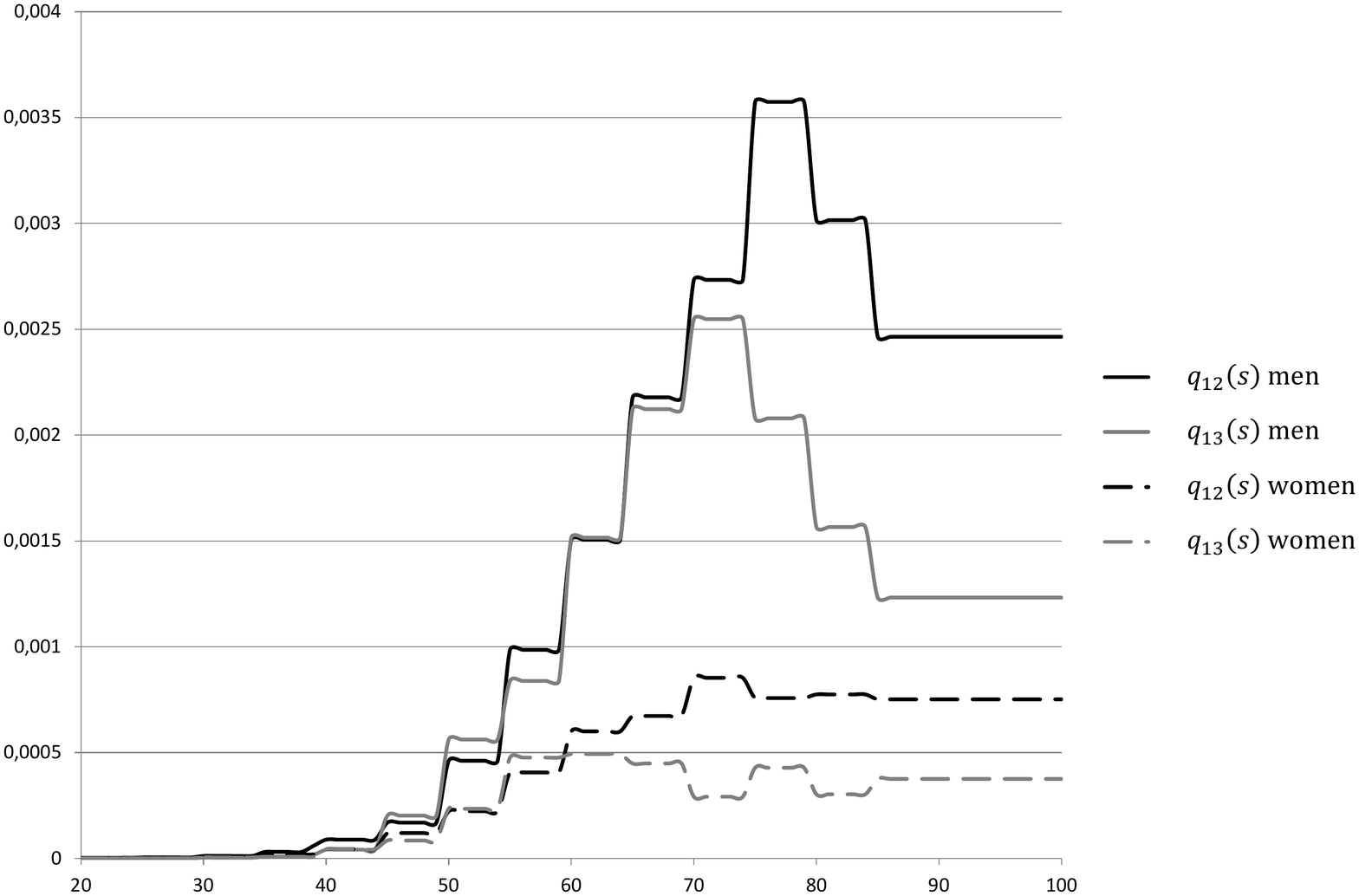}
\caption{\label{Fig.3}  Transition probability from state
1 to state 2 or 3.}
\end{figure}

In Figure~\ref{Fig.3} we present probabilities \refs{est_q_12} and \refs{est_q_13}
with respect to age and sex of an insured person.
Looking at the graphs, a significant
difference in incidence between men and women can be observed. Men
succumb to lung cancer several times more often than women. A
different regularity is also noticeable. In the older age groups
diagnosis without metastases is often posed. For women it is
around 60 year of age, for men around 70. In the youngest groups,
lung cancer is diagnosed with metastases. In interpreting the
results, it should be remembered that the disease is considered
from the point of view of a calendar year, rather than the annual
individual patient's medical history. Thus, a history of an
insured person who enters the oncological health care system in
January (at the beginning of a year) looks quite different from
that of a patient who was diagnosed in December (at the end of a
year). For this reason, the percentage of people diagnosed with
metastases may appear to be lower than it might seem from the
epidemiological data. The chosen method of analysis enables to
take into account the fact that an insured person could
potentially fall ill throughout the year since the inception of
the insurance contract. In the event that this happens at the
beginning of this year, the chance of occurrence of metastases is
higher than when it happens at the end of the year.

\begin{figure}[h!bt]
\hspace{2cm}
\includegraphics[width=11cm]{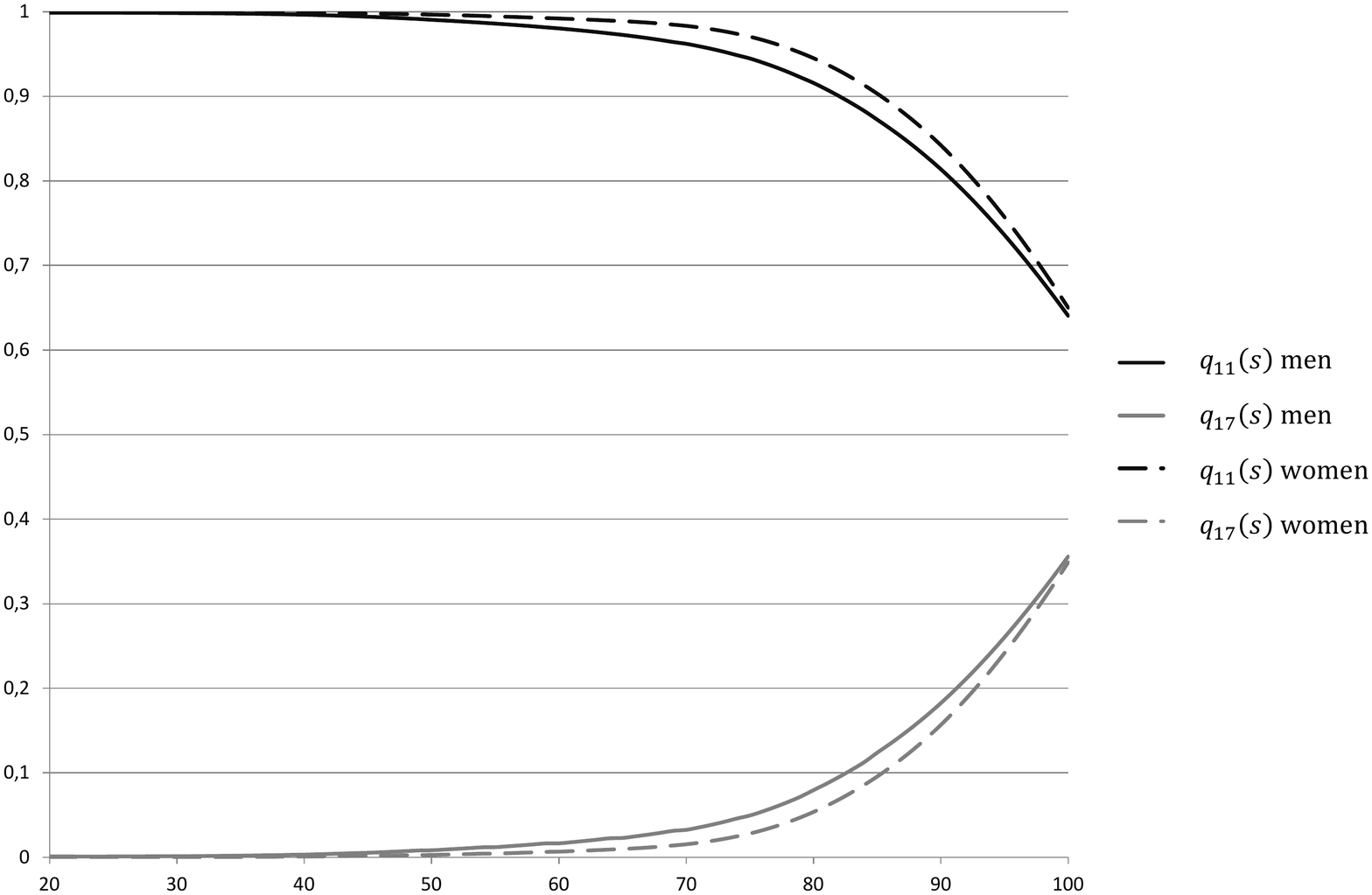}
\caption{\label{Fig.4} Transition probability from state 1 to
state 7, and the probability of remaining in state 1.}
\end{figure}

Based on Polish Life Tables 2008 \cite{TTZ-GUS},
the probabilities \refs{est_q_17} and \refs{est_q_11}
with respect to age and sex of a person are presented in
Figure~\ref{Fig.4}.

\subsection{Lung cancer without metastasis}

Evaluating of transition probabilities for patients who were
diagnosed with cancer without metastases (i.e. probabilities of
the second row of the matrix \refs{macierzQ}) is the next step of
the analysis. Estimating of probabilities is associated with the
analysis of the history of  hospitalization of patients with lung
cancer who, during the first admission in 2008, had no metastasis.
Patients were hospitalized for the first time in 2006, 2007 or
2008.  There is a need to define the cohort of patients, who were
ill in 2006 and 2007 and at the beginning of 2008 did not have
metastases as well as patients, who fell ill in 2008 without
metastases as the first diagnosis. The percentage of patient who,
during this year metastases were diagnosed, is calculated. The
transition probability $q_{23}(k)$ can be estimated using
$\varrho_{x+k}$, which is the proportion of patients suffering
from lung cancer in 2008, who got metastases during the year and
it is estimate based on \cite{NHF}. In further considerations, we
accept that $q_{27}(k)=q_{x+k}$.
It is connected with the fact, that a sick person,
who has not metastases, has a higher risk of death than a healthy
person, so we also take into account the possibility of dying for
one of many reasons, including lung cancer. Taking into account
the above considerations, we obtain $q_{22}(k)=
1-q_{x+k}-\varrho_{x+k}$.

Patients who, during 2008, suffered from lung cancer and in the
initial diagnosis in 2008 they did not have metastases, pose the
studied cohort at this stage. 1098 men and 533 women belonged to
the analyzed cohorts.

In case of the second state, the examination of history of the
disease from the perspective of an insurance company makes the
chance of an insured person to remain without metastasis
apparently higher than it results from the epidemiological data.
While comparing the empirical percentages of diagnosed metastases
calculated for particular years of age in a given year in
populations of men and women, some differences can be spotted for
both populations. In women population, the percentage of diagnosis
with metastasis is highest in the age group of 45 years, then it
gradually decreases. In male population it grows, reaching a peak
in the age group about of 60 years, then subsides gradually. It
should be noted that, in age groups with the highest lung cancer
incidence rates (from 50 to 70 years), the probability of
diagnosis of metastasis for a patient with determined lung cancer
is higher in men population.

The chance of getting a diagnosis of metastatic is modelled using
a Bernoulli distributed random variable.
The parameter $\varrho$ is defined as the success probability and is estimated
using the logistic regression model.
In this model, an independent variable
is an age of a patient. In both populations, patients below 45
years of age occur very rarely. For this group of patients the
probability of diagnosis with metastases is calculated using the
nearest neighbor method. This means that the probability is
constant in this group and equals the probability of diagnosis for
a 45 years-old person. In female population, logistic regression
model parameters were estimated for patients over 45 years of age.
In the population of men, patients above 45 years of age are
divided into two groups, the first form 45 to 59 years of age, and
the second above 59 years of age. In these two groups, the
probabilities of diagnosis with metastases are estimated using
separate logistic regression models. A specified age group
division ensures the best fitting of the model to the data.

Parameter estimators of models for male and female populations are
presented in Table~\ref{Tab.3}.
\begin{table}[h!bt]
 \begin{center}
  \caption{Parameter estimators}
{\small
\begin{tabular}{|l|c|c|c|}  \hline \label{Tab.3}
Parameter                                       & Estimator &
Stand. error & p-value
\\ \hline
In female population (above 45 years of age)    &           &
&
\\
Age (in years)                                  &-0.024468& 0.001822 & 0.00000 \\
\hline
In male population (from 45 to 59 years of age) &         &          &          \\
Constant                                        &-6.27958 &2.231157  & 0.004885 \\
Age (in years)                                  & 0.09215 &0.040526  & 0.023000 \\
\hline
In male population (from above 59 years of age) &         &          &          \\
Constant                                        & 3.447079&1.173173  & 0.003301 \\
Age (in years)                                  &-0.074952&0.017151  & 0.000012 \\
\hline
\end{tabular}}
 \end{center}
\end{table}

The goodness of fit of model is evaluated on the basis of Wald's
test results, Lemesow Hosmer test and values of deviation.
Selection of model was based on information criterion AIC. The
results are presented in Table~\ref{Tab.4}.

\begin{table}[h!bt]
 \begin{center}
  \caption{The goodness of fit of models}
{\small
\begin{tabular}{|l|c|c|c|} \hline
\multicolumn{4}{|l|}{For female population (above 45 years of
age)} \\ \hline \label{Tab.4} Results of Hosmer Lemesow Test  &
{Test statistics}  & p-value    &  \\ \hline
                                & 7.4058            & 0.3878     &  \\ \hline
Statistic of goodness of fit    & Degree of freedom & Statistics &
Stat/Df   \\ \hline Deviation                       & 520 &
476.550    & 0.916443  \\ \hline AIC &                   & 478.550
&           \\ \hline \multicolumn{4}{|l|}{For male population
(from 45 to 59 years of age)}       \\ \hline {Result of Wald's
test}          & Test statistics   & p-value    &           \\
\hline Constant & 7.921377          & 0.004885   &           \\
\hline Age (in years)                  & 5.168502          &
0.023000   &
\\ \hline Statistic of goodness of fit    & Degree of freedom &
Statistics & Stat/Df   \\ \hline Deviation                       &
329               & 346.231    & 1.052374  \\ \hline AIC
&                   & 350.231    &           \\ \hline Results of
Hosmer Lemesow Test  & Test statistics   & p-value    &
\\ \hline
                                & 4.7341            & 0.4491     &           \\ \hline
\multicolumn{4}{|l|}{For male population (above 59 years of age)}
\\ \hline Result of Wald's test           & Test statistics   &
p-value    &           \\ \hline Constant                        &
8.63333           & 0.003301   &           \\ \hline Age (in
years)                  & 19.09870          & 0.000012   &
\\ \hline Statistic of goodness of fit    & Degree of freedom &
Statistics & Stat/Df   \\ \hline Deviation                       &
747               & 611.179    & 0.818178  \\ \hline AIC
&                   & 615.179    &           \\ \hline Results of
Hosmer Lemesow Test  & Test statistics   & p-value    &
\\ \hline
                                & 7.2045            & 0.3023     &           \\ \hline

\end{tabular}}
 \end{center}
\end{table}

\begin{figure}[h!bt]
\hspace{2cm}
\includegraphics[width=11cm]{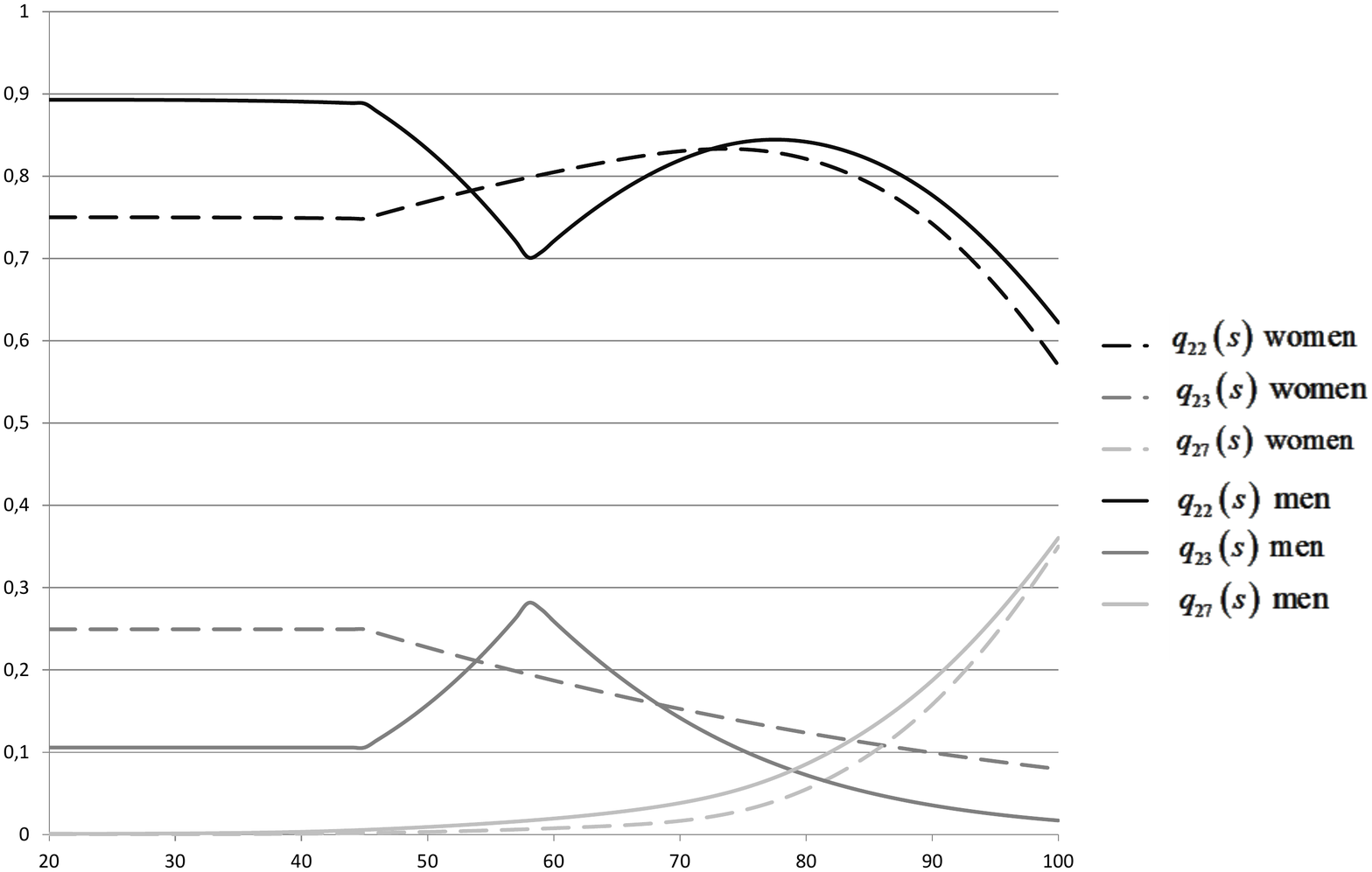}
\caption{\label{Fig.5} Probabilities connected with state 2.}
\end{figure}

Transition probabilities associated with state 2 are presented in Figure~\ref{Fig.5}.
All the needed probabilities $q_{x+k}$ were taken from Polish Life Tables 2008 \cite{TTZ-GUS}.

\subsection{Lung cancer with metastasis}\label{sec:with_metastasis}

After receiving the diagnosis of metastasis, a patient is
considered to be terminally ill. Note that all state connected
with this situation (i.e. $i=3,4,5,6$) are reflex states, then
$q_{ii}(k)=0$ and we get
\begin{eqnarray}
q_{i8}(k)&=& 1-q_{i i+1}(k). \label{q_i9}
\end{eqnarray}
A person with diagnosed metastases lives no longer than four
years, so $q_{68}(k)=1$.

In particular, the estimation of $q_{34}(k)$ (and
$q_{38}(k)$) is equivalent to modelling of the survival time
of patients, who had metastatic disease before 2008, who were
diagnosed with metastatic during their first visit in 2008 or they
got metastases during 2008.

Due to the low incidence rate of lung cancer with metastatic among young
people between 20 and 39 years old, the probabilities associated with the
states from 3 to 6 were estimated using the nearest neighbor method.
Hence this probabilities are equal to probabilities calculated for
40-year old persons.

Based on analysis of the
mortality of the cohort members at age $s=40,41,...,100$, we
estimate transition probabilities \refs{q_i9} for $i=3,4,5$.
Let us introduce a variable $T_s$, which denotes the number of
years that the $s$-years-old patient from the analyzed cohort
survived. On the basis of empirical data \cite{NHF} we assume that
$T_s$ takes values from $0$ to $3$. If $T_s=0$, which means that a
patient died during the first year (counting from the first
hospitalization with diagnosed metastases during 2008). If
$T_s=1$, it means that a patient died during the second year et
cetera. A patient survives for maximum four years.

Male population consists of 845 patients. The
empirical distribution of the number of survived years is
presented in Table~\ref{Tab.5}.
\begin{table}[h!bt]
 \begin{center}
  \caption{The empirical distribution of number of survived years for men.}
{\small
\begin{tabular}{|c|c|c|} \hline \label{Tab.5}
Number of survived years & Number of patients  & Empirical
distribution \\ \hline
0 & 732 & 0.8662722 \\ \hline
1 & 84  & 0.0994083 \\ \hline
2 & 27  & 0.0319527 \\ \hline
3 & 2   & 0.0023668 \\ \hline
\end{tabular}}
 \end{center}
\end{table}

In male population, variable $T_s$ is modelled using logistic
regression for ordered categorical variable. The estimators of
parameters are presented in Table~\ref{Tab.6} point A. The
significance of regression coefficients was verified on the basis
of Wald test results (p-value$< 0.01$). The results of test are
shown in Table~\ref{Tab.6} point B. The goodness of fit model was
verified on the basis of Akaike criterion (AIC), deviation and
value of chi-square statistic, which are shown in
Table~\ref{Tab.6} point C.
\begin{table}[h!bt]
 \begin{center}
  \caption{Modelling $T_s$ for men.}
{\small
\begin{tabular}{|l|c|c|c|} \hline
\multicolumn{4}{|l|}{A. Parameter estimators} \\ \hline
\label{Tab.6} Parameter                     & estimator   &
p-value   &   \\ \hline
Constant3                     & 3.208851 &  0.001690 &   \\ \hline
Age (in years)                & 0.044698 &  0.000168  &   \\ \hline
\multicolumn{4}{|l|}{B. The results of Wald test} \\ \hline
Parameter & Wald Statistic      & p-value   &   \\ \hline
Constant    &  96.27373           & 0.000000  &    \\ \hline
Age         &  14.15864           & 0.000168  &     \\ \hline
\multicolumn{4}{|l|}{C. The statistics goodness of fit} \\ \hline
                      & Degree of freedoom & Value of Statistic & Stat/Df \\ \hline
Deviation             & 2531   & 793.688   & 0.313587 \\ \hline
Chi-square statistic  & 2531   & 2193.335  & 0.866588  \\ \hline
AIC                   &        & 801.688   &           \\ \hline
\end{tabular}}
 \end{center}
\end{table}

The age of a patient occurred a significant factor which has an
influence on survival. Average influence of age is expressed by
the slope of age. Additionally, we observed that age of a patient
determines significantly the probability of survival for two and
three years. This fact is reflected by a significant estimator of
constant 3. On the basis of the model, the following probabilities
can be estimated:
\begin{eqnarray} \label{ZEqnNum141703}
\Prob\left(T_s \le 2\right)=\frac{\exp
\left(3.208851+0.044698s\right)}{1+\exp
\left(3.208851+0.044698s\right)}
\end{eqnarray}
and
\begin{eqnarray} \label{ZEqnNum473129}
\Prob\left(T_s \le 1\right)=\frac{\exp \left(0.044698
s\right)}{1+\exp \left(0.044698 s \right)} ,
\end{eqnarray}
where $s$ denotes the age of a patient. Taking into account that
survival is equal to maximum 3, we obtain
$\Prob\left(T_s=3\right)=1-\Prob\left(T_s \le 2\right).$ On the
basis of \refs{ZEqnNum141703} and \refs{ZEqnNum473129}, we
calculate $\Prob\left(T_s =2\right)$.
Due to the fact that age affected the survival for one
and two years on average quite similarly, probabilities
$\Prob\left(T_s=0\right)$ and $\Prob\left(T_s=1\right)$are
calculated as weighted probabilities of \refs{ZEqnNum473129} in
the following way
\begin{eqnarray} \label{1.16)}
\Prob\left(T_s=0\right)=w_{0} \cdot \Prob\left(T_s\le 1\right),\;
\Prob\left(T_s=1\right)=w_{1} \cdot \Prob\left(T_s\le 1\right).
\end{eqnarray}
The weights $w_{0} =0.897059$ and $w_{1} =1-w_{0} =0.102941$
denote the percentage of patients, who do not survive one year and
the percentage of those that survived one year, respectively
in the group of all patients who died within one and two years.
They are estimated for the whole cohort without dividing into age
groups. This is because the estimated probability
$\Prob\left(T_s\le 1\right)$ takes into account the average effect
of age on survival in the entire cohort. On the basis of the
probability distribution of variable $T_s$, the transition
probabilities connected with state $3$ are calculated as
${q_{38}(s) =\Prob\left(T_s=0\right)}$ and
${q_{34}(s) =\Prob\left(T_s=1\right)+\Prob\left(T_s=2\right)+\Prob\left(T_s=3\right).}$
On the basis of conditional probabilities, we estimate for
$i=4,5,6$ the transition probabilities to state $8$ in the
following way
$q_{i8}(s) =\Prob\left(T_s=i-3\left|T_s>i-4\right.
\right)=\frac{\Prob\left(T_s=i-3\right)}{1-\sum_{j=0}^{i-4}\Prob\left(T_s=j\right)}$.
Finally we obtain
\begin{eqnarray}
q_{38}(s)&=& \left\{
\begin{array}{lll}
 0.768485 &{\rm for}& s \in [20,40] \\
 0.89706  m(s) &{\rm for}& s > 40
\end{array} \label{q38men}
\right. , \\
q_{48}(s)&=& \left\{
\begin{array}{lll}
 0.380912 &{\rm for}& s \in [20,40] \\
 \frac{0.10294 m(s)}{1-0.89706 m(s)} &{\rm for}& s > 40
\end{array}
\right. , \\
q_{58}(s)&=& \left\{
\begin{array}{lll}
 0.953154 &{\rm for}& s \in [20,40] \\
 \frac{\frac{\exp(3.20885+0.044698s)}{1+\exp(3.20885+0.044698s)}-m(s)}{m(s)} &{\rm for}& s > 40
\end{array} \label{q58men}
\right. ,
\end{eqnarray}
where $m(s)=\frac{ \exp(0.044698s)}{1+\exp(0.044698s)}$.

Note that according to \refs{q_i9}, the survival probabilities
connected with the third to sixth states are defined by the
probabilities of death.

In female population, 324 patients belong to the cohort, which is
analyzed in the third state. The empirical distribution of the number of
survived years is presented in Table~\ref{Tab.9}.

\begin{table}[h!bt]
 \begin{center}
  \caption{The empirical distribution of number of survived years for women.}
{\small
\begin{tabular}{|c|c|c|} \hline \label{Tab.9}
Number of survived years& Number of patients & Empirical
distribution           \\ \hline 0                       & 268 &
0.8271605 \\ \hline 1                       & 43  & 0.1327160 \\
\hline 2                       & 12  & 0.0370370\\ \hline 3
& 1   & 0.0030865 \\ \hline
\end{tabular}}
 \end{center}
\end{table}

In case of female population, the Poisson regression with identity
link function is used to model the probability of survival. The
estimators of parameters are presented in Table~\ref{Tab.10} point
A. The significance of regression coefficients was verified on the
basis of Wald test results (p-value$<0.01$), compare
Table~\ref{Tab.10} point B. Then the goodness of fit of model was
verified on the basis of Akaike criterion (AIC) and deviation and
value of chi-square statistic, which are shown in
Table~\ref{Tab.10} point C.

\begin{table}[h!bt]
 \begin{center}
  \caption{Modelling $T_s$ for women.}
{\small
\begin{tabular}{|l|c|c|c|} \hline
\multicolumn{4}{|l|}{A. Parameter estimators} \\ \hline
\label{Tab.10} Parameter                     & estimator     &
p-value     &    \\ \hline Constant                      &
0.552179      &  0.000606   &    \\ \hline Age (in years)
& -0.005435     & 0.025888    &    \\ \hline
\multicolumn{4}{|l|}{B. The results of Wald test} \\ \hline
Parameter                     & Wald Statistic    & p-value    &
\\ \hline Constant                      & 10.01119          &
0.001556   &  \\ \hline Age                           &  4.22647
& 0.039798   &  \\ \hline \multicolumn{4}{|l|}{C. The statistics
goodness of fit} \\ \hline
                      & Degree of freedoom & Value of Statistic & Stat/Df \\ \hline
Deviation             & 322                & 246.934           &
0.766875 \\ \hline Chi-square statistic  & 322                &
378.150            & 1.174379  \\ \hline AIC                   &
& 370.912           &           \\ \hline
\end{tabular}}
 \end{center}
\end{table}

The age of a patient turned out to be a significant factor
influencing survival chances. The probability of surviving
\textit{k} years is calculated using the following formula
\begin{eqnarray} \label{1.21)}
\Prob\left(T_s=k\right)=\frac{\lambda^k}{k!} \exp
\left(-\lambda\right)
\end{eqnarray}
where $\lambda =\E\left(T_s\right)=-0.005435s+0.552179$ and
$k=0,1,2$. The probability that a patient survives over two years
is expressed by $\Prob\left(T_s=3\right)=1-\sum
_{k=0}^{2}\Prob\left(T_s=k\right) $.

The probabilities of death $q_{i8}(k)$ (for $i = 3,4,5$)
are calculated similarly to those for men and we obtain
\begin{eqnarray}
q_{38}(s)&=& \left\{
\begin{array}{lll}
 0.715503 &{\rm for}& s \in [20,40] \\
 \exp(-w(s))&{\rm for}& s > 40
\end{array} \label{q38women}
\right. , \\
q_{48}(s)&=& \left\{
\begin{array}{lll}
 0.841937 &{\rm for}& s \in [20,40] \\
 \frac{w(s)\exp(-w(s))}{1-\exp(-w(s))} &{\rm for}& s > 40
\end{array}
\right. , \\
q_{58}(s)&=& \left\{
\begin{array}{lll}
 0.891591 &{\rm for}& s \in [20,40] \\
 \frac{0.5w(s)^2\exp(-w(s))}{1-(1+w(s))\exp(-w(s))}&{\rm for}& s > 40
\end{array} \label{q58women}
\right. ,
\end{eqnarray}
where $w(s)=-0.005435s+0.552179$.

\begin{figure}[h!bt]
\hspace{0.5cm}
\includegraphics[width=14cm]{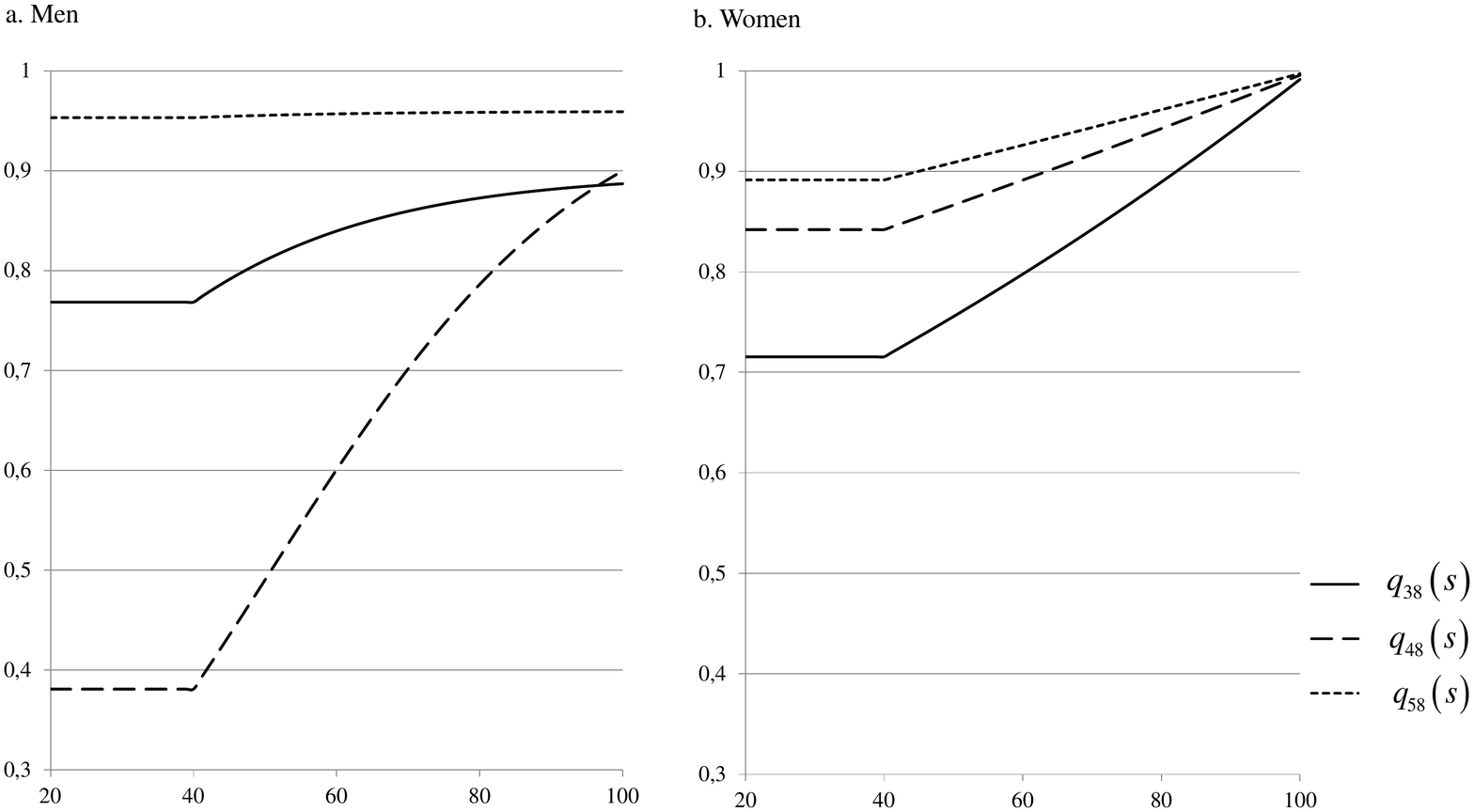}
\caption{\label{Fig.6} Probabilities of death for terminally ill persons.}
\end{figure}

In Figure~\ref{Fig.6} we present probabilities of death for terminally ill men (Figure~\ref{Fig.6}a) and women (Figure~\ref{Fig.6}b) with respect to age.

The statistical analyzes presented in Section \ref{sec:E.p.t} are
carried out by means of Statistica~10.

\section{Conclusions}\label{sec:conclusion}

The multiple state model presented in Section~\ref{sec:mm} is
suitable not only for modelling the critical illness insurance
contracts but also for other health insurances. It can also be
adapted to insurance contracts against the loss of income due to
disability or the loss of health (income protection). The
introduced model allows for combining CII with life insurance.
In such a combination disease benefits are usually provided
as an acceleration benefit to a life insurance.

The results of
Sections~\ref{sec:prob.structure}-\ref{sec:E.p.t} can be directly used
to build the multiple increment-decrement tables for (proposed in Section~\ref{sec:mm})
the multiple state model connected with lung cancer in the following form
\begin{eqnarray}
\left\{q_{11}(s),q_{12}(s),q_{13}(s),q_{22}(s),q_{23}(s),q_{34}(s),q_{45}(s),q_{56}(s)
\right\}_{s \geq 0}. \nonumber
\end{eqnarray}
Such tables are useful for the valuation of
insurance contracts (premiums and reserves) or outflows from
Health Found consisting of health-related expenses.

Death probabilities \refs{q38men}-\refs{q58men} and \refs{q38women}-\refs{q58women} concerning the population of those suffering from lung
cancer with metastasis (e.g. states $3, 4, 5, 6, 8$) are needed
to calculate the value of viatical settlement payments.

\end{document}